\documentclass[twocolumn,twoside,slac_two]{revtex4}
\usepackage{graphicx}
\usepackage{fancyhdr}
\input babarsym

\begin{document}

\title{Rare $\tau$ decays}


\author{Stefano Passaggio}
\affiliation{INFN, Sezione di Genova, via Dodecaneso 33, 16146 Genova, Italy}

\begin{abstract}
A review is presented of the current status of experimental searches for physics beyond 
the standard model in rare $\tau$ decays.
\end{abstract}

\maketitle

\thispagestyle{fancy}


\section{Introduction}
The availability of large samples of $\tau$ leptons allows for searches
of very rare (Standard Model forbidden) $\tau$ decay modes, which provide 
a sensitive probe for physics
beyond the Standard Model (BSM). In recent years, very large samples of $\tau$ 
leptons have been made available by the high luminosity running of the
two asymmetric $B$-factories
$\pep2$ and KEK $B$, which, in light of the comparable size of the 
$\epem \rightarrow \bbbar$ and $\epem \rightarrow \tautau$ cross sections
($\sigma_{\tautau} \sim 0.9 \nb$, at $\sqrt{s} \sim M_{\Upsilon(4S)}$), 
can rightfully be regarded as $\tau$ factories as well. In fact, the 
luminosity so far recorded by the experiments $\babar$ and Belle running
respectively at $\pep2$ and KEK $B$ (${\cal L}_{\babar} \sim 470 \invfb$, and 
${\cal L}_{Belle} \sim 710 \invfb$) entails that the total sample of $\tau$ pairs
produced in the two experiments is already currently in excess of $10^9$.

Differently from the quark sector, the lepton world is well known to
be characterized by an apparent remarkable conservation of flavor. An exact
conservation of Lepton Flavor (LF) is naturally implemented in the Standard
Model through the assumption that neutrinos are mass degenerate (actually: massless) 
fermions. 
Since the observation of neutrino oscillations, we know that the 
latter assumption is actually wrong, and that, with it, we also have to give
up with the strict conservation of LF. 
\begin{figure}
\includegraphics{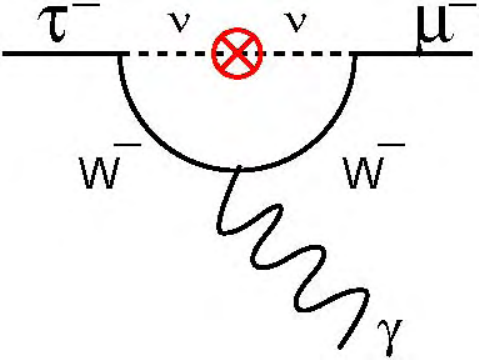}
\caption{Dominant SM diagram contributing to the radiative LFV decay 
         $\tau \rightarrow \mu\gamma$.\label{fig:SM1}}
\end{figure}
However, the minimal extension 
of the SM that amounts to simply allowing for massive and nondegenerate
neutrinos\footnote{I will henceforth refer to such ``na\"ive'' minimal extension of the 
Standard Model as ``SM'', without further comments upon it actually being an extension
of the Standard Model proper. It amounts to simply extending the Standard Model by the 
addition of an appropriate neutrino mass matrix in a gauge invariant manner, while keeping
the minimal Higgs scheme, with only one Higgs doublet.}, due to the remarkable effectiveness 
of the ensuing leptonic GIM 
mechanism which in turn stems from the smallness of the neutrino masses,
predicts extremely low rates for all LF violating (LFV) phenomena in the charged 
lepton sector, notwithstanding the current evidence in favor of large
leptonic mixing. Using the largest value in the 
currently favored experimental range~\cite{PDG07} 
for $\Delta m^2_{32} \equiv \left| m^2_3 - m^2_2 \right| \sim 3 \ 10^{-3} \ev^2$, 
where $m_i, \; i=1,2,3$ are the three neutrino mass eigenvalues, and allowing 
for maximal mixing, the SM expected rate~\cite{LeeShrock77,ChengLi80} for the radiative 
LFV decay $\tau \rightarrow \mu\gamma$ (see Fig.~\ref{fig:SM1} for the dominant SM diagram)
amounts to a surely unobservable $\BR(\tau^- \rightarrow \mu^- \gamma) = 
\frac{3\alpha}{128\pi}\left(\frac{\Delta m^2_{32}}{M^2_W}\right)^2 \sin^2 2 \theta_{23}
\;\BR(\tau^- \rightarrow \mu^- \overline{\nu}_{\mu} \nu_{\tau}) \sim 10^{-54}$.

\begin{figure}
\includegraphics{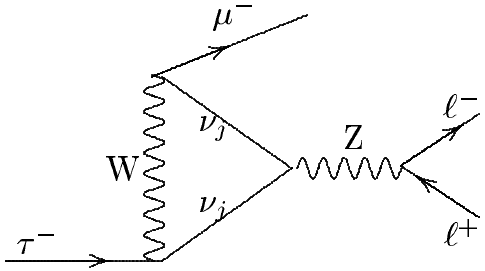}
\caption{One loop SM diagram contributing to the nonradiative LFV decay 
         $\tau^- \rightarrow \mu^-l^+l^-$ (from~\cite{Pham99}).\label{fig:Pham1}}
\end{figure}

It is worthwhile to remark that the SM suppression of LFV processes in the charged sector,
though a general feature of the model, doesn't 
achieve to the same degree, and in particular to the extreme one mentioned above, for all processes.
As noted by Pham~\cite{Pham99}, LFV $\tau$ decays of the kind of $\tau^- \rightarrow \mu^-l^+l^-$ or
$\tau^- \rightarrow \mu^- \rho^0$ ($l = e, \; \mu$), due to the presence of infrared
divergences in the contributing diagram shown in Fig.~\ref{fig:Pham1}, 
get in fact a milder, logarithmic, suppression factor 
in place of the striking factor $\left(\Delta m^2_{ij}/M^2_W\right)^2$ suppressing the class of
radiative LFV processes $\tau \rightarrow l \gamma$. 
The SM prediction for the rate of such nonradiative LFV processes turns in fact to be proportional to
$\sum^{3}_{k=2} U_{\tau k} U^{\star}_{\mu k} \ln \frac{m^2_i}{m^2_1}$. Taking into account 
the current experimental knowledge on neutrino mixing, this translates to expected branching ratios 
many orders of magnitude larger than those expected for for $\tau \rightarrow l \gamma$
(Pham~\cite{Pham99} estimates $\BR \gtrsim 10^{-14}$), but in any case far below 
the current experimental reach.

The above considerations allow us to draw two important conclusions.
The first one is that, although the discovery that neutrinos are massive entails that
LF numbers are no longer conserved in the SM,
we can safely state that, as long as the SM represents
the correct physical description, no LFV
process in the charged sector (CLFV) should ever be observed. This means that observation of 
CLFV would undoubtedly be a signature of BSM physics. 
The second conclusion is that it is worthwhile and important to search for
CLFV in all possible experimentally searchable processes (e.g. radiative
and nonradiative ones), since: (a) as shown to be the case in the SM, also in
most BSM physics scenarios different CLFV processes could be characterized
by widely different rates, and (b) the eventual observation and measurement of LFV in different
processes can be generally expected to be sensitive to different sectors of the new physics parameter
space and better constraining the choice of a new scenario. 

In fact, many scenarios of BSM physics predict rates for LFV $\tau$ decays that
are not too far, or even within, the current experimental reach.
Without pretending to be exhaustive, table~\ref{Tab:BSMscenarios} shows some examples 
of scenarios that have been theoretically investigated, with the corresponding
order of magnitude expectations for the branching ratio of radiative and nonradiative 
LFV $\tau$ decays.
\begin{table}[h]
\begin{center}
\caption{Examples of BSM scenarios predicting rates of LFV $\tau$ decays not too far, or even
within, the current experimental reach. Order of magnitude expectations for the two broad classes
of radiative (e.g. $\tau \rightarrow l \gamma$) and nonradiative (e.g. $\tau \rightarrow 3l $) LFV
$\tau$ decays are displayed in the last two columns.}
\begin{tabular}{|c|c|c|}
\hline 
 & \textbf{$\BR (\tau \rightarrow l \gamma)$} & \textbf{$\BR (\tau \rightarrow 3l)$} \\
\hline
 SUSY Higgs~\cite{DedesEtAl02,BrignoleRossi03}             & $10^{-10}$ & $10^{-7}$  \\
\hline
 SM + Heavy Majorana $\nu_R$~\cite{CveticEtAl02}           & $10^{-9}$  & $10^{-10}$ \\
\hline
 Non-universal $Z^{\prime}$~\cite{YueEtAl02}               & $10^{-9}$  & $10^{-8}$  \\
\hline
 SUSY $SO(10)$~\cite{MasieroEtAl03,FukuyamaEtAl03}         & $10^{-8}$  & $10^{-10}$ \\
\hline
 mSUGRA + Seesaw~\cite{EllisGomezEtAl02,EllisHisanoEtAl02} & $10^{-7}$  & $10^{-9}$  \\
\hline
\end{tabular}
\label{Tab:BSMscenarios}
\end{center}
\end{table}

\section{Searching for LFV $\tau$ decays at $\epem$ colliders}
The $\babar$~\cite{BaBarDet} and Belle~\cite{BelleDet} detectors are remarkably similar, with the major difference being in
the technology used to identify charged particles: Belle uses a threshold \v{C}erenkov detector
together with time-of-flight and tracker dE/dx, whereas $\babar$ mainly relies on a ring-imaging
\v{C}erenkov detector augmented by dE/dx in the trackers.

\begin{figure}
\includegraphics{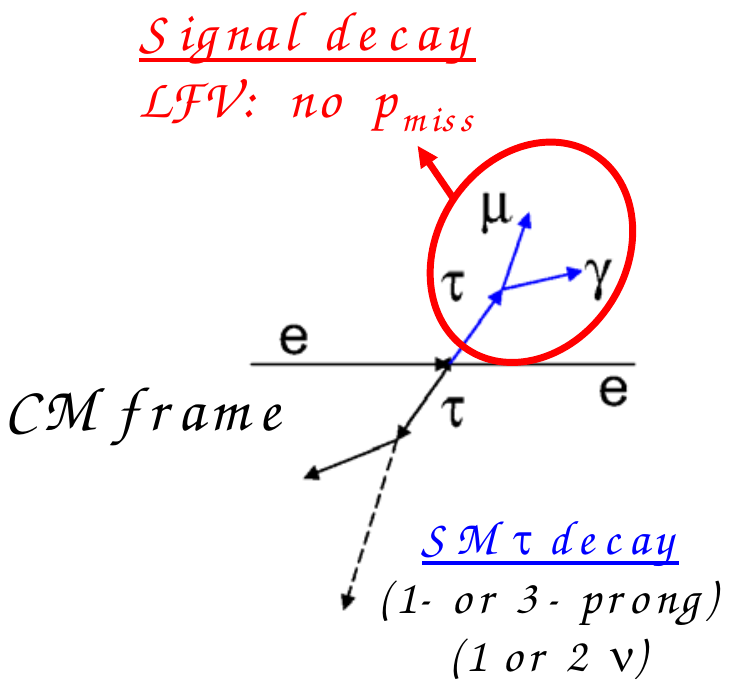}
\caption{Schematic drawing of the production of $\tau$ pairs in $\epem$ collisions, displayed in the CM frame.
         The figure also illustrates the general strategy employed in searches for LFV $\tau$ decays at
         $\epem$ colliders: one $\tau$ is reconstructed in a SM allowed (``Tag'') decay, 
         characterized by a 1- or 3-prong + 1 or 2 $\nu$ final state, while the search for a possible LFV
         (neutrinoless!) decay is performed on the recoiling (``Signal'') $\tau$.\label{fig:eeToTauTauCM}}
\end{figure}

In addition to the statistics issue already discussed in the introduction, 
$\epem$ colliders running at, or nearby, the $\Upsilon(4S)$ resonance provide
a very clean environment for the search of rare (and, in particular, LFV) $\tau$ decays.
$\tau$ leptons are exclusively\footnote{Apart from possible initial or final state radiation $\gamma$('s).} 
produced in pairs through the QED process $\epem \rightarrow \tautau$, and thereby
fly back-to-back in the $\epem$ Center of Mass (CM) frame (Fig.~\ref{fig:eeToTauTauCM}).
At the asymmetric $B$-factories $\pep2$ and KEK $B$ the events are boosted in the laboratory (Lab) frame with a 
$\beta\gamma \simeq 0.56$ for the $\babar$ experiment and $\simeq 0.43$ for Belle.
%, as illustratedin Fig.~\ref{fig:MCeeToTauTauLab}. 
The boost of each of the two $\tau$'s in the CM frame, combined with such a
CM-to-Lab boost, confer to the final state of each $\tau$ decay a ``jet-like'' shape which allows not only
to powerfully reject the $\BB$ background, but also to easily separate the final state products of one $\tau$
from those of the other. 
%\begin{turnpage}
%\begin{figure*}
%\includegraphics[width=135mm]{figure4.pdf}
%\caption{Example of a simulated $\epem \rightarrow \tautau$, with the ``Signal'' (in this example: negative) $\tau$ 
%         decaying as $\tau^- \rightarrow \mu^- \gamma)$ and the ``Tag'' $\tau$ decaying to the 1-prong + 2 $\nu$'s
%         final state $\ep \nu_e \overline{\nu}_{\tau}$, seen in laboratory frame of the $\babar$ 
%         experiment.\label{fig:MCeeToTauTauLab}}
%\end{figure*}
%\end{turnpage}

The general strategy employed in searches for LFV $\tau$ decays at
$\epem$ colliders consists in searching for $\epem \rightarrow \tautau$ events where one $\tau$ is reconstructed 
in a SM allowed decay (henceforth referred to as ``Tag'' decay), characterized by a 1- or 3-prong + 1 or 2 $\nu$ 
final state, while the search for a possible LFV decay (LFVD) is performed on the recoiling (``Signal'') $\tau$.
A noteworthy feature of all LFV $\tau$ decays searched for by $\babar$ and Belle is that such decays do not
bear any undetectable particle (specifically: any neutrino) in their final state, and can therefore be completely
reconstructed. In other words, the observed missing momentum of the reconstructed complete event is wholly attributable
to the ``Tag'' decay. This circumstance provides very powerful tools for background rejection, which amount to
requesting that the reconstructed invariant mass and CM energy for the candidate LFVD (``Signal'') side of the event
coincide with the nominal mass of the $\tau$ lepton and with half the CM energy of the $\epem$ collision, respectively:
\begin{equation}
M_{\mathrm{LFVD}} = m_{\tau}
\label{eq:M}
\end{equation}
\begin{equation}
\Delta E \equiv E^{CM}_{\mathrm{LFVD}} - \frac{\sqrt{s}}{2} = 0 
\label{eq:DeltaE}
\end{equation}

\begin{figure}%[hbt]
\includegraphics[width=80mm]{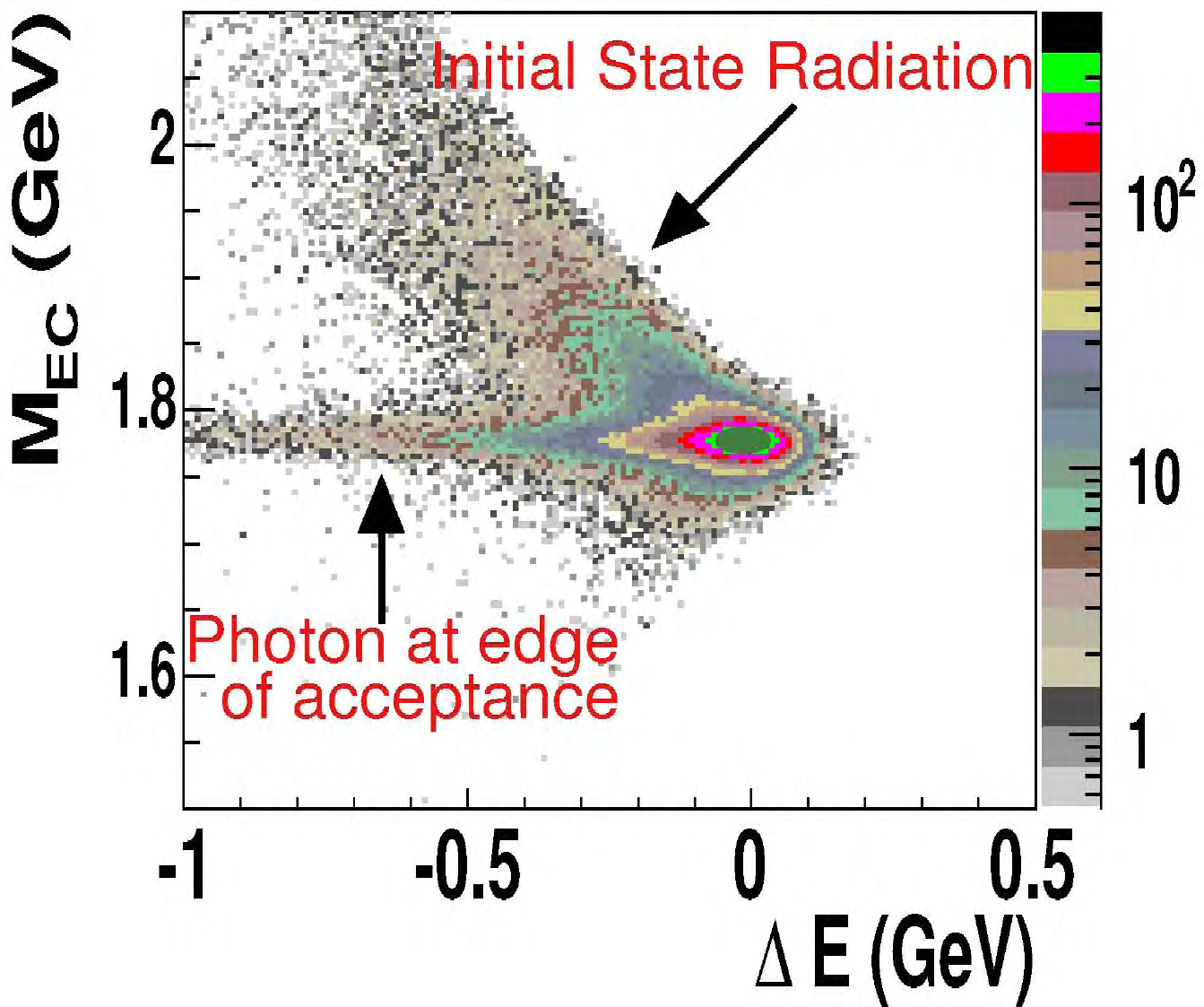}
\caption{Scatter plot of the reconstructed variables $M_{\mathrm{LFVD}}$ and $\Delta E$ for 
         MC simulated $\epem \rightarrow \tautau$ events where one of the $\tau$ leptons decays to the LFV 
         final state $\mu \gamma$ (from the $\babar$ experiment).
         For the exact meaning of the variable plotted along the vertical axis ($M_{\mathrm{EC}}$), see the text and
         footnote~\ref{ftn:MEC}. 
         The dark green ellipse drawn on top of the distribution in correspondence of the constraints expressed by
         Eqs.~\ref{eq:M} and \ref{eq:DeltaE} illustrates in a qualitative fashion the relative size
         of a typical $2 - 3$ standard deviations wide signal box employed in the final event selection.
         \label{fig:MC_DeltaEvsM_taumugamma}}
\end{figure}

An example of the distribution of simulated signal ($\tau \rightarrow \mu \gamma$) MC events in a scatter plot 
of these two kinematical variables is shown in Fig.~\ref{fig:MC_DeltaEvsM_taumugamma}, which displays the accumulation
of signal events in correspondence of the two constraints given by Eqs.~\ref{eq:M} and \ref{eq:DeltaE}, as
well as the smearing brought about by resolution and radiative effects, respectively.
The same figure displays also in a qualitative fashion
a two dimensional signal box which is used to separate the signal from the SM backgrounds.

Exploiting at best the available constraints can significantly boost the sensitivity of the search. An example
of this important remark is offered by the strategy employed by the $\babar$ experiment since the publication
of their search for $\tau \rightarrow \mu \gamma$ decays~\cite{BaBar_TauMuGamma}. The choice of evaluating
$M_{\mathrm{LFVD}}$ by imposing a beam energy constraint\footnote{Whence the use of the symbol $M_{\mathrm{EC}}$ in 
the scatter plot of Fig.~\ref{fig:MC_DeltaEvsM_taumugamma}. \label{ftn:MEC}}, obtained by fixing the ``Signal'' $\tau$ decay vertex -- or equivalently the emission point of the photon in the final state -- in correspondence of the point
of closest approach of the signal $\mu$ trajectory to the beam axis, brought a dramatical improvement in terms
of resolution on such a kinematical variable\footnote{The resolution improvement on $M_{\mathrm{LFVD}}$
due to the beam energy constraint technique ($M_{\mathrm{EC}}$) is greater of a factor two: 
$\sigma(M_{\mathrm{EC}}) \sim 9 \mev$, while $\sigma(M_{\mathrm{inv}}) \sim 20 \mev$, where $M_{\mathrm{inv}}$ denotes
the straightforward evaluation of the kinematical quantity $M_{\mathrm{LFVD}}$ as an unconstrained invariant mass of 
the LFVD decay products.} with respect to a more straightforward reconstruction
of the same quantity as an unconstrained invariant mass of the LFVD decay products. Typical values for the resolution
on the other kinematical quantity $\Delta E$ range around $\sim 50 \mev$.
\begin{figure}[bt]
\includegraphics[width=80mm]{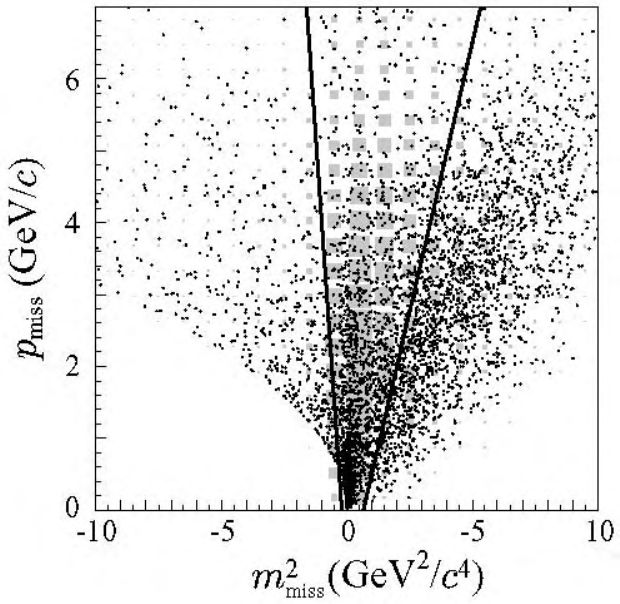}
\caption{Scatter plot of the reconstructed variables $p_{\mathrm{miss}}$ and $m^2_{\mathrm{miss}}$ for
         data (dots) and signal MC (shaded boxes) events in a search for the LFV decay $\tau \rightarrow \mu \gamma$
         by the Belle collaboration ~\cite{Belle_TauLGamma}. The selection criterion employed consists in requiring that
         the event lies between the two lines in the figure.
         \label{fig:MCandData_pmissVsMmisssq}}
\end{figure}

Other ingredients crucial to an effective reduction of backgrounds consist in a clever use of the powerful particle
identification capabilities of the two experiments and, whether possible, in the exploitation of additional constraints
on the event. An example of the latter strategy is represented by the requirement that the reconstructed event missing mass
($m_{\mathrm{miss}}$),
when the ``Tag'' decay is chosen so as to have a single neutrino in the final state, be compatible with zero. Such a constraint
proved to be very effective in Belle's searches for the LFV decay 
$\tau \rightarrow \mu \gamma$~\cite{Belle_TauLGamma}, as
illustrated in Fig.~\ref{fig:MCandData_pmissVsMmisssq}, where the selection criterion on $m^2_{\mathrm{miss}}$ employed 
by the Belle collaboration
is seen to be implemented in a $p_{\mathrm{miss}}$-dependent fashion.

\begin{figure}
\includegraphics[width=80mm]{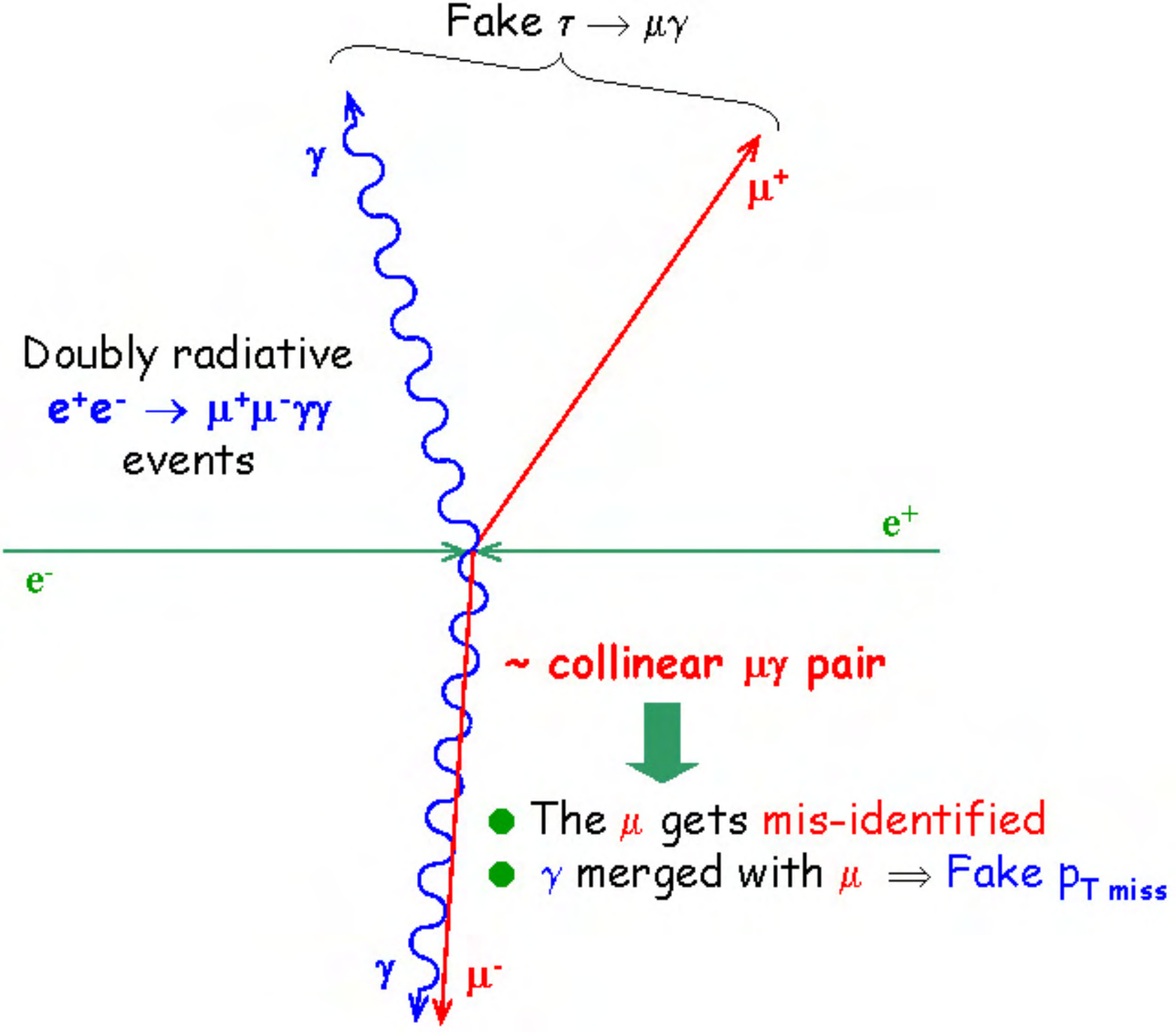}
\caption{Schematic representation of a doubly radiative, $\epem \rightarrow \mumu\gamma\gamma$, fake missing momentum event.
         \label{fig:FakePmiss}}
\end{figure}

In addition, a rewarding strategy has in some cases turned out to consist in trying to identify, and then cleverly fight, specific, particularly nasty, backgrounds which could otherwise survive all other selection criteria. An example of this
situation is represented by a fraction of the residual $\mu\mu\gamma(\mathrm{s})$ background in searches 
for the $\tau \rightarrow \mu \gamma$ decay, namely that consisting of doubly
(or more) radiative events where the flight path of one of the photons lies sufficiently close to the trajectory of the charged track in the ``Tag'' side of the event. In these occurrences, schematically illustrated in Fig.~\ref{fig:FakePmiss},
such a photon may fail to be reconstructed as a neutral cluster in the electromagnetic calorimeter and
its energy deposit in the calorimeter could eventually get associated to the nearby charged track.
When the photon is sufficiently energetic, such occurrences would mimic a substantial missing
momentum. Concurrently, the ``Tag'' side charged track would erroneously fail to be identified
as a muon according to basically any of the standard $\mu$ identification criteria because of the
large energy deposit in the calorimeter erroneously associated to it.

As a general rule, both $\babar$ and Belle optimize\footnote{The analyses are optimized to give the best 
``expected upper limit''.} their selection strategy, and the particular values of the associated
cuts, by using samples of fully simulated and reconstructed signal and background MC events. MC samples, 
possibly checked with control samples or with data events sufficiently far from the signal region (sidebands), are also 
used to model the background shapes of relevant kinematical variables for the purpose of estimating
and subtracting the final background contribution, whose normalization is obtained from sidebands data. 
The MC simulation of the signal is also used to
determine the signal efficiency ($\epsilon$), which typically lies between $2\%$ and $10\%$, depending on
the channel under study.
To minimize possible biases, both experiments
adopt a blind analysis approach, which consists in 
excluding from consideration, up to the moment when the full selection and all systematic studies are completely finalized, all events in the data within a suitably shaped and sized region
in the $M_{\mathrm{LFVD}}$ and $\Delta E$ plane around the signal signature values given by Eqs.~\ref{eq:M} and \ref{eq:DeltaE}.

In order to possibly boost the sensitivity of the search, the two experiments have adopted a spectrum of different
approaches to finally discriminate a possible signal from the residual background and obtain an estimate of the
former's size: these range from simple cut-and-count to likelihood techniques, and in one case a Neural-Network
selection has been adopted. If the estimated background  ($N_{\mathrm{bkd}}$) is compatible with the observed 
number of events, a $90\%$ confidence level (CL) upper limit on the number of signal events is estimated\footnote{Also
in this respect, the two experiments have resorted to a number of different techniques~\cite{Narsky00,Junk99andRead02}. 
It should be noted
that the experimental upper limits quoted by both the experiments are in any case of a frequentistic 
nature. Negative fluctuations are generally admitted (and occur!) and give rise to ``measured'' upper limits
that can be substantially lower than the ``sensitivity'' or ``expected upper limit'' of the corresponding search.
\label{ftn:freq}} ($N^{UL}_{90}$). Based on this quantity and on the estimate of the signal efficiency $\epsilon$,
a $90\%$ CL upper limit on the branching ratio is obtained as:
\begin{equation}
\BR^{UL}_{90} = \frac{N^{UL}_{90}}{2 {\cal L} \sigma_{\tautau} \epsilon}.
\label{eq:BRUL}
\end{equation}

\section{Experimental results}
Both $\babar$ and Belle have searched for LFV in many different classes of $\tau$ decays. In the following
I'll summarize the results obtained by the two experiments, grouping them according to the decay final state
searched for and giving particular emphasis only to the most recent ones.

\subsection{Search for $\tau \rightarrow l \gamma$ ($l=e$, $\mu$)}
\begin{figure}
\includegraphics[width=80mm]{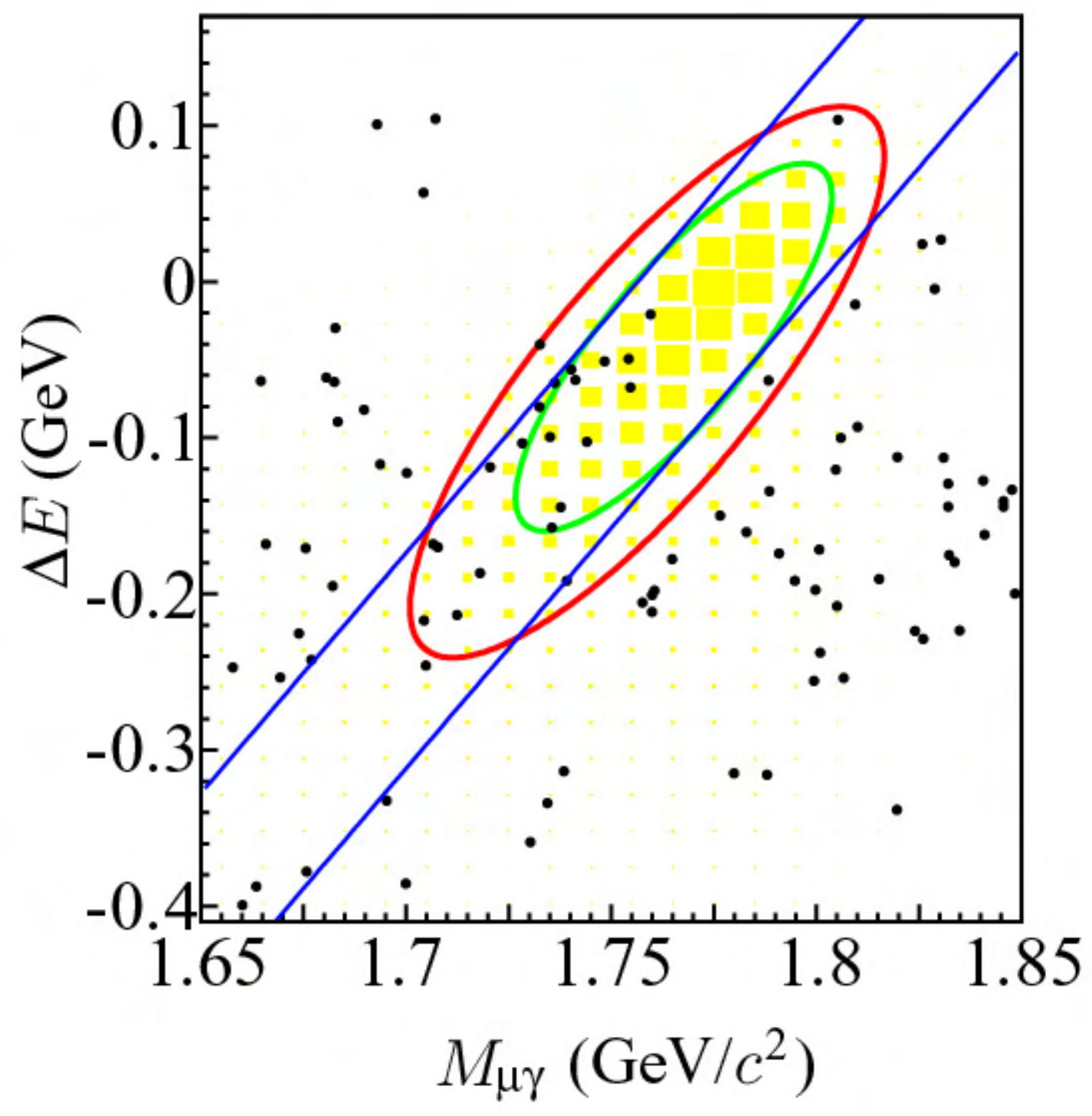}
\caption{Scatter plot of the reconstructed variables $\Delta E$ and $M_{\mathrm{LFVD}}$ for the final event selection
         obtained in the most recent Belle search for $\tau \rightarrow \mu \gamma$. Dots are data events and yellow squares
         represent MC signal events. The inner green ellipse is the $2\sigma$ signal region (where a total of 10 data events
         are observed, with an estimated efficiency $\epsilon = 5.1\%$), while the outer red ellipse is the $3\sigma$ 
         blinded region.
         \label{fig:BelleTauMuGamma}}
\end{figure}

\begin{figure}
\includegraphics[width=80mm]{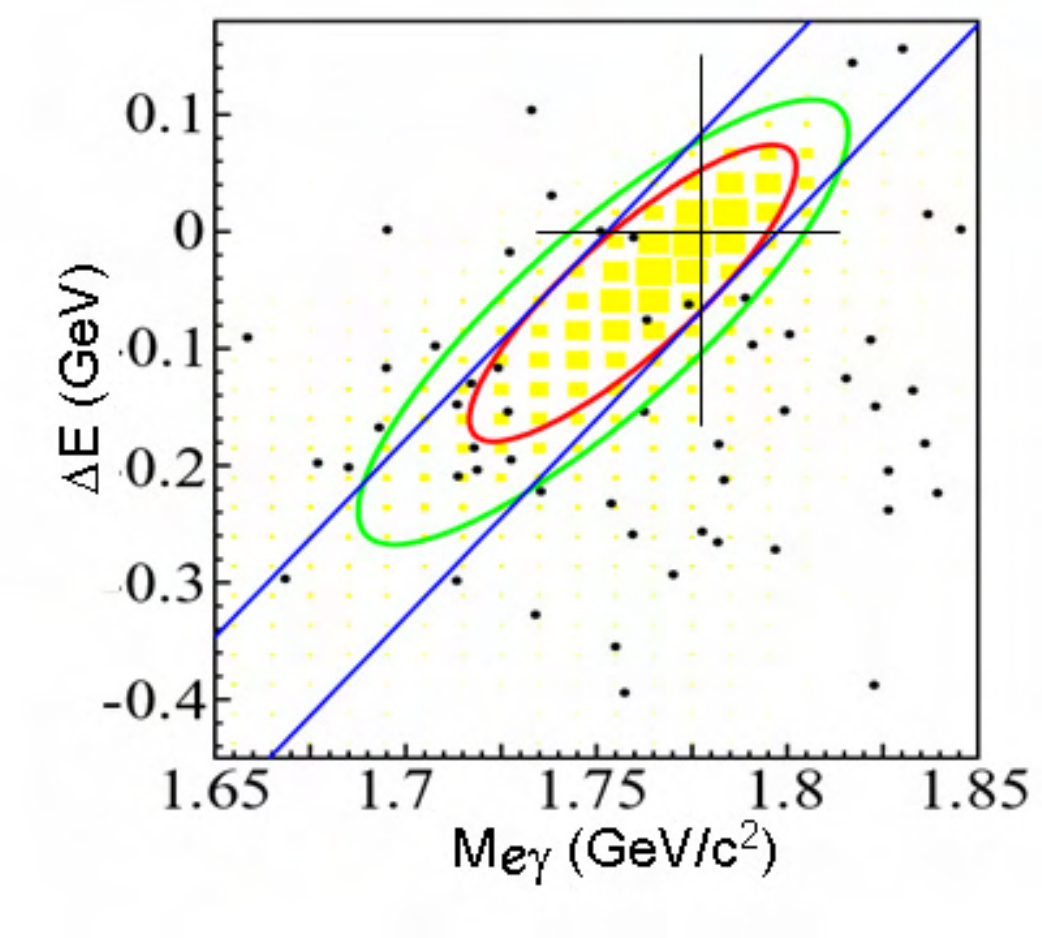}
\caption{Scatter plot of the reconstructed variables $\Delta E$ and $M_{\mathrm{LFVD}}$ for the final event selection
         obtained in the most recent Belle search for $\tau \rightarrow e \gamma$. Dots are data events and yellow squares
         represent MC signal events. The inner red ellipse is the $2\sigma$ signal region (where a total of 5 data events
         are observed, with an estimated efficiency $\epsilon = 3.0\%$), while the outer green ellipse is the $3\sigma$ 
         blinded region.
         \label{fig:BelleTauEGamma}}
\end{figure}

The most recent $\tau \rightarrow \mu \gamma$ and $\tau \rightarrow e \gamma$ results were 
submitted
for publication by Belle~\cite{Belle_TauLGamma} using a data sample corresponding to a luminosity ${\cal L} 
= 535 \invfb$.
The $\tau \rightarrow \mu \gamma$ ($\tau \rightarrow e \gamma$) analyses have a $5.1\%$ ($3.0\%$) signal efficiency within a 2D elliptical signal
region in the $M_{\mathrm{LFVD}}$ and $\Delta E$ plane (see Fig.~\ref{fig:BelleTauMuGamma} and
Fig.~\ref{fig:BelleTauEGamma}). 

After performing a 2D unbinned extended maximum likelihood
fit for the number of signal ($s$) and background ($b$) events, Belle finds $s = -3.9^{+3.6}_{-3.2}$ ($-0.14^{+2.18}_{-2.45}$) 
and $b = 13.9^{+6.0}_{-4.8}$ ($5.14^{+3.86}_{-2.81}$)
in the $\tau \rightarrow \mu \gamma$ ($\tau \rightarrow e \gamma$) search. 
Toy MC simulations are then used to evaluate the probability of obtaining such results\footnote{Belle evaluates that 
$P(s \leq -3.9) = 25\%$ ($P(s \leq -0.14) = 48\%$) for null true signal in the $\tau \rightarrow \mu \gamma$ 
($\tau \rightarrow e \gamma$) search.} and to estimate the $90\%$ CL
upper limits on the number of signal events: $N^{UL}_{90}(\tau \rightarrow \mu \gamma) = 2.0$ and 
$N^{UL}_{90}(\tau \rightarrow e \gamma) = 3.3$.
The branching ratio upper limits following from these results amount respectively to
$\BR(\tau \rightarrow \mu \gamma) < 4.5 \; 10^{-8}$ and $\BR(\tau \rightarrow e \gamma) < 1.2 \; 10^{-7}$ at $90\%$ CL.

The corresponding limits previously obtained by $\babar$ on ${\cal L} = 211 \invfb$ amount to 
$\BR(\tau \rightarrow \mu \gamma) < 6.8 \; 10^{-8}$~\cite{BaBar_TauMuGamma} and 
$\BR(\tau \rightarrow e \gamma) < 1.1 \; 10^{-7}$~\cite{BaBar_TauEGamma} at $90\%$ CL.

\begin{figure*}%[h]
\includegraphics[width=170mm]{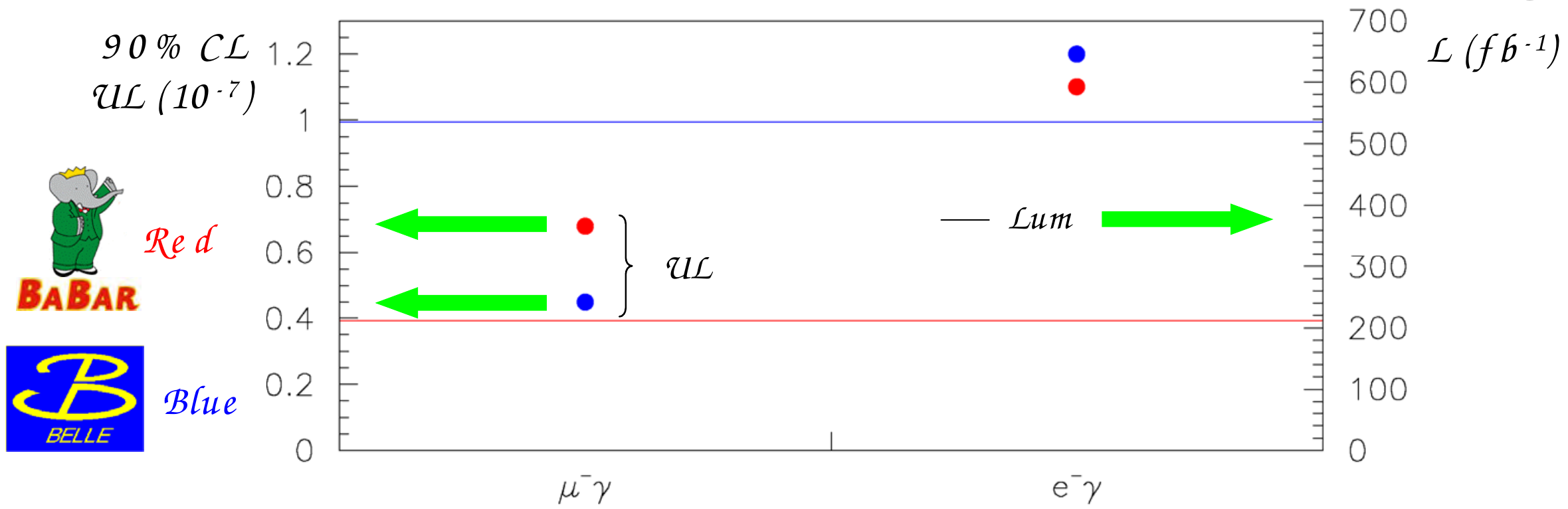}
\caption{Summary and comparison of the current upper limits (dots: left vertical axis) 
         set by $\babar$ (red) and Belle (blue) 
         on $\BR(\tau \rightarrow \mu \gamma)$ and $\BR(\tau \rightarrow e \gamma)$. The corresponding analyzed luminosities
         are shown as horizontal lines on a scale displayed on the right vertical axis.
         \label{fig:BaBarBelleTauLGamma}}
\end{figure*}

Fig.~\ref{fig:BaBarBelleTauLGamma} summarizes and compares the $\babar$ and Belle results discussed above, and the
corresponding analyzed luminosities.

\subsection{Search for $\tau \rightarrow l \pi^0$, $l \eta$, $l \eta^{\prime}$, $l K^0_S$}
\begin{figure*}%[h]
\includegraphics[width=160mm]{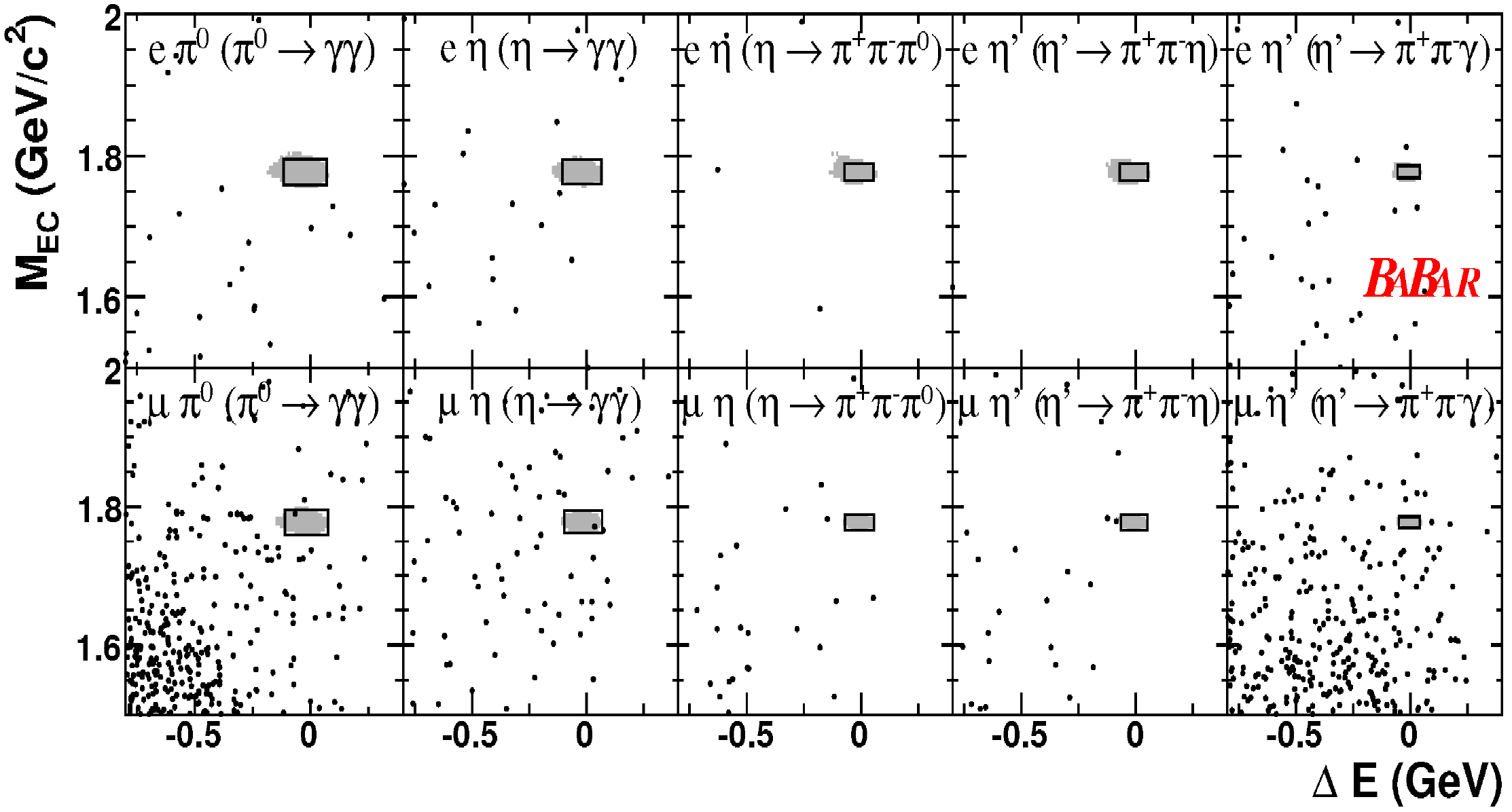}
\caption{Selected data (dots) and $68\%$
         of signal MC events (shaded region) in the $M_{\mathrm{LFVD}}$ and $\Delta E$ plane for the 10 LFV decay channels
         $\tau \rightarrow l \pi^0$, $l \eta$, $l \eta^{\prime}$ searched for by $\babar$; for each channel, the
         corresponing $2\sigma$ signal box is displayed.
         \label{fig:BaBarLPS}}
\end{figure*}

Both $\babar$~\cite{BaBar_TauLPS} and Belle~\cite{BelleLPS1} have recently published 
new results on LFV $\tau$ decays involving the $\pi^0$, $\eta$ and $\eta^{\prime}$ 
pseudoscalars: $\tau \rightarrow l \pi^0$, $l \eta$, $l \eta^{\prime}$, where $l$ is separately identified as either an electron or a muon. The luminosity analyzed by $\babar$ is ${\cal L} = 339 \invfb$, while Belle used 
${\cal L} = 401 \invfb$.
In the modes with an $\eta$ meson in the final state both the $\eta \rightarrow \gamma\gamma$ and the 
$\eta \rightarrow \pi^+ \pi^- \pi^0$ decays are used. In the $\tau \rightarrow l \eta^{\prime}$ analyses the 
$\eta^{\prime} \rightarrow \pi^+ \pi^- \eta$ ($\eta \rightarrow \gamma\gamma$) and $\eta^{\prime} \rightarrow \rho^0 \gamma$ 
decay modes were included. Fig.~\ref{fig:BaBarLPS} shows the distribution of the finally selected data events in the
$M_{\mathrm{LFVD}}$ and $\Delta E$ plane for each of these channels, together with the $2\sigma$ signal box used in the $\babar$ search.

The $\babar$ expected background per channel is between 0.1 and 0.3 events. Summing over all ten
modes, the total expected background within the signal regions amounts to 3.1 events, whereas 2 events in
total were observed.

Belle had previously published a search for an additional mode with a neutral pseudoscalar meson in the final state 
($\tau \rightarrow l K^0_S$~\cite{BelleLPS2}, ${\cal L} = 281 \invfb$).

\begin{figure*}%[h]
\includegraphics[width=170mm]{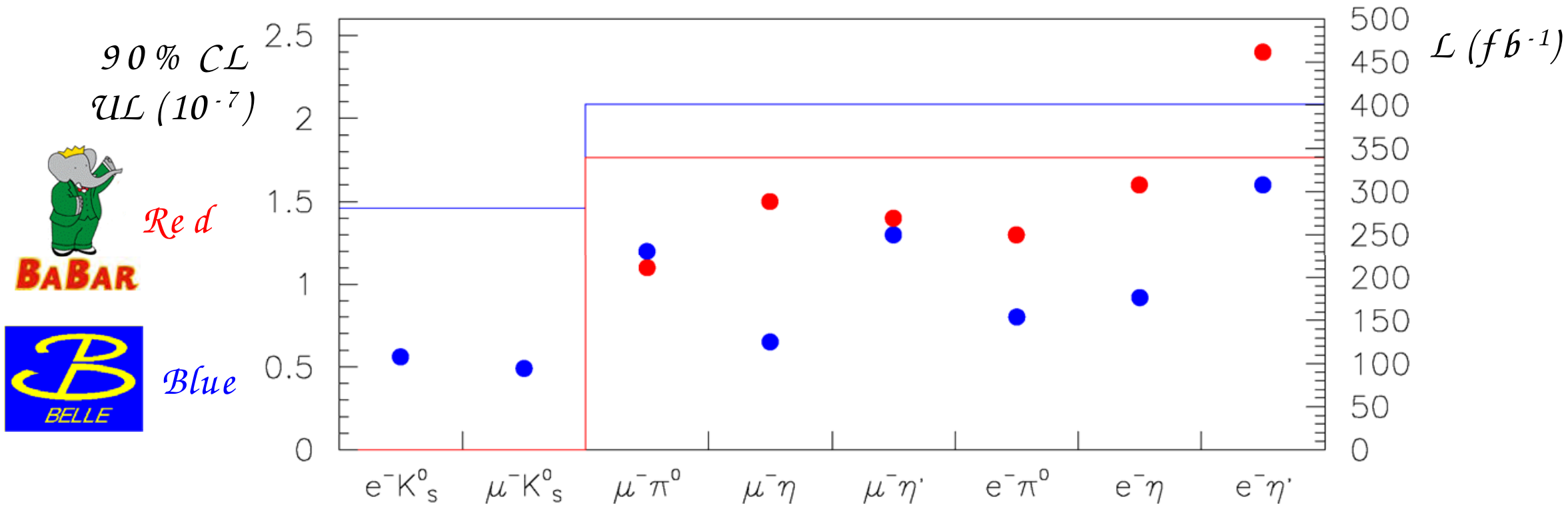}
\caption{Summary and comparison of the current upper limits (dots: left vertical axis) 
         set by $\babar$ (red) and Belle (blue) 
         on the branching ratios for $\tau$ decays to a lepton and a neutral pseudoscalar meson. 
         The corresponding analyzed luminosities
         are shown as horizontal lines on a scale displayed on the right vertical axis.
         \label{fig:BaBarBelleTauLPS}}
\end{figure*}

Fig.~\ref{fig:BaBarBelleTauLPS} summarizes and compares the $\babar$ and Belle results on all searches
for $\tau$ LFV decays to a lepton and a neutral pseudoscalar meson.

\subsection{Other searches for $\tau$ LFV decays}
\begin{figure*}%[h]
\includegraphics[width=170mm]{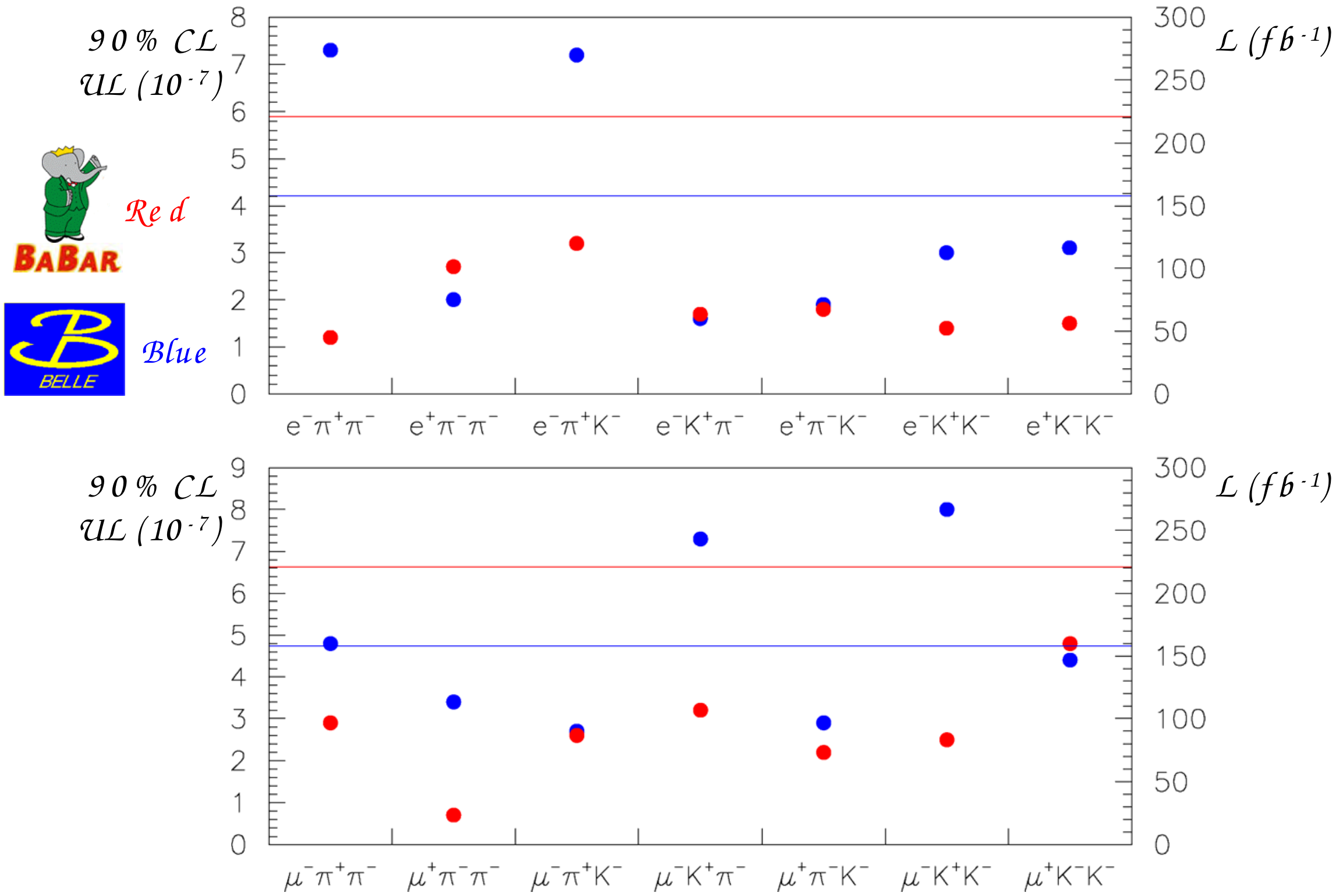}
\caption{Summary and comparison of the current upper limits (dots: left vertical axis) 
         set by $\babar$~\cite{BaBar_TauLhh} (red) and Belle~\cite{Belle_TauLhhAndTauLV0} (blue) 
         on the branching ratios for $\tau$ decays to a lepton and a pair of charged pseudoscalar mesons. 
         The corresponding analyzed luminosities
         are shown as horizontal lines on a scale displayed on the right vertical axis.
         \label{fig:BaBarBelleTauLhh}}
\end{figure*}

\begin{figure*}%[h]
\includegraphics[width=170mm]{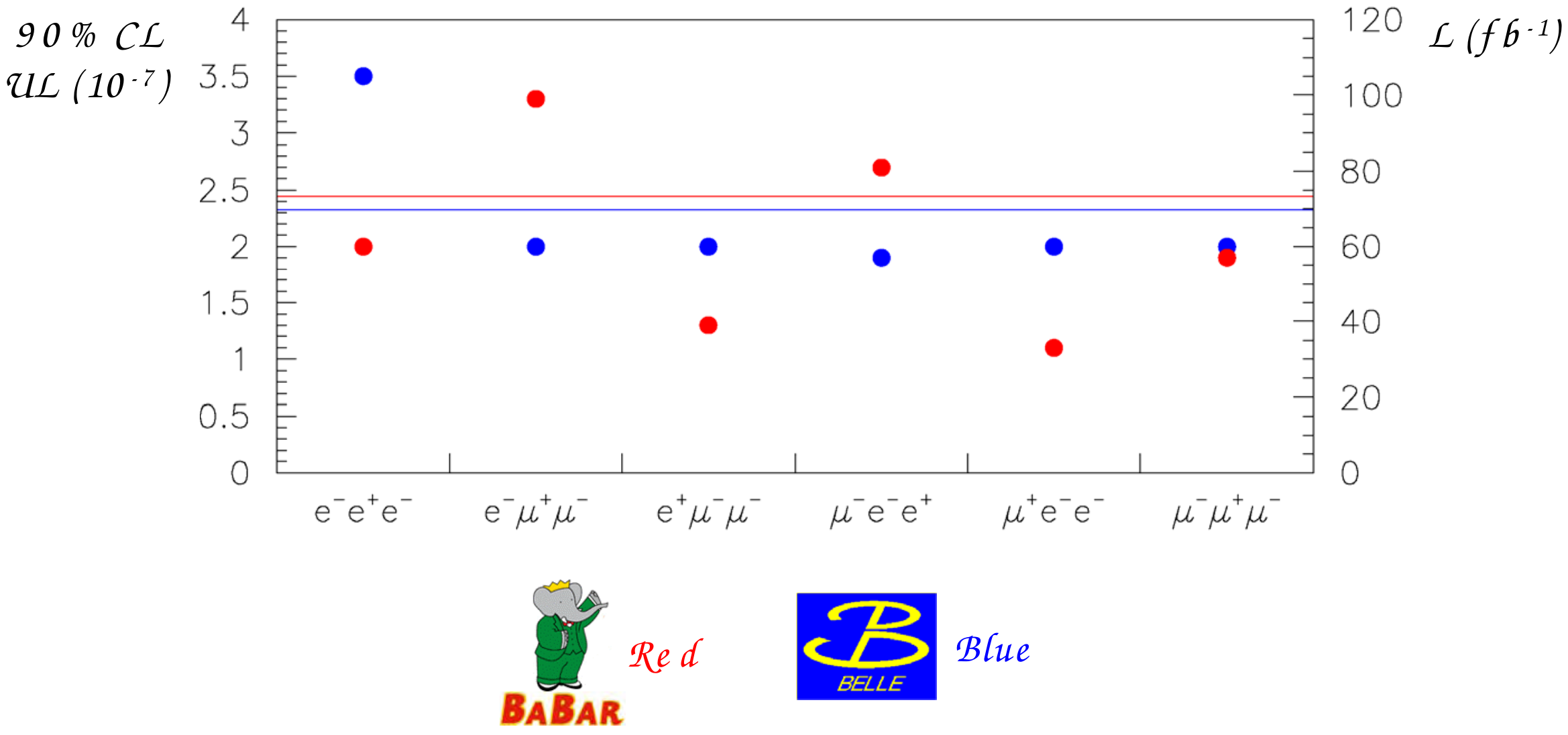}
\caption{Summary and comparison of the current published upper limits (dots: left vertical axis) 
         set by $\babar$~\cite{Babar_TauLll} (red) and Belle~\cite{Belle_TauLll} (blue) 
         on the branching ratios for $\tau$ decays to three charged leptons. 
         The corresponding analyzed luminosities
         are shown as horizontal lines on a scale displayed on the right vertical axis.
         \label{fig:BaBarBelleTauLll}}
\end{figure*}

\begin{figure*}%[h]
\includegraphics[width=170mm]{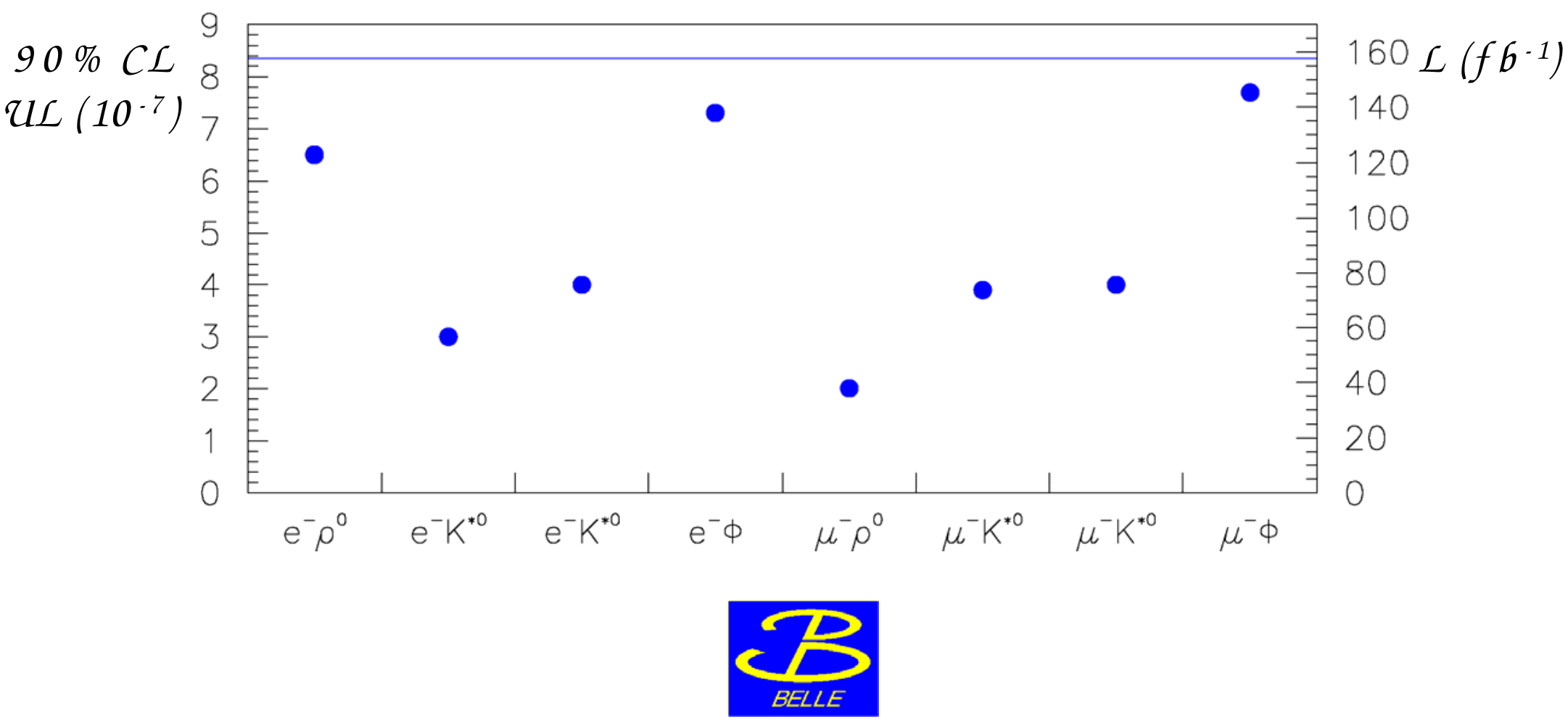}
\caption{Summary of the current upper limits (dots: left vertical axis) 
         set by Belle~\cite{Belle_TauLhhAndTauLV0} 
         on the branching ratios for $\tau$ decays to a lepton and a neutral vector meson. 
         The corresponding analyzed luminosity
         is shown as a horizontal line on the scale displayed on the right vertical axis.
         \label{fig:BelleTauLV0}}
\end{figure*}

Both $\babar$ and Belle have searched for other $\tau$ LFV decays. The corresponding results are compared
in Fig.~\ref{fig:BaBarBelleTauLhh} for the class of decays $\tau \rightarrow l h h^{\prime}$ (where
$h$ and $h^{\prime}$ $= \pi^{\pm}$ or $K^{\pm}$)~\cite{BaBar_TauLhh,Belle_TauLhhAndTauLV0}. Fig.~\ref{fig:BaBarBelleTauLll} compares the results
obtained by the two experiments for the class of decays 
$\tau \rightarrow 3l$~\cite{Babar_TauLll,Belle_TauLll}\footnote{$\babar$ has very recently
submitted for publication~\cite{BaBar_TauLll2} 
an updated search for $\tau \rightarrow 3l$, using a much larger luminosity 
(${\cal L} = 376 \invfb$) than previously published by each of the two experiments (see Fig.~\ref{fig:BaBarBelleTauLll}). 
Upper limits 
on the branching fractions are set in the range $(4-8) \; 10^{-8}$ at $90\%$ CL. These results were not public
when FPCP2007 took place, and were not presented at the Conference.}
, while Fig.~\ref{fig:BelleTauLV0}
shows the results published by Belle on several $\tau$ decays to a lepton and a neutral vector
meson~\cite{Belle_TauLhhAndTauLV0}.

\section{Future Prospects for LFV searches in $\tau$ decays}
The estimated physics reach of the full data sample that will be collected by $\babar$ and Belle
till the end of their foreseen running, based on projections from existing analyses, depends on
the residual background level. We express the experimental reach in terms of the ``expected
$90\%$ CL upper limit'' and, for brevity's sake, refer to this as the ``sensitivity''. In the absence of
signal, for large $N_{\mathrm{bkd}}$, $N^{UL}_{90} \sim 1.64 N_{\mathrm{bkd}}$, whereas for small $N_{\mathrm{bkd}}$ 
a value for $N^{UL}_{90}$ is obtained
using the method described in~\cite{CousinsHighland92}. So, for $N_{\mathrm{bkd}} \sim 0$, $N^{UL}_{90} \sim 2.4$.
Reducing the background below
a handful of events doesn't greatly improve the expected limit if significant efficiency is lost in
the process, which is why it is common to see experiments reporting the expected backgrounds
to be small (i.e. a few events), but rarely below 0.1 of an event.

A worst-case scenario is obtained if identical analyses to those already published by $\babar$ and
Belle are repeated, as is, on the increased data sample: in that case the expectations then simply
scale as $\sqrt{N_{\mathrm{bkd}}}/{\cal L}$, which, for large $N_{\mathrm{bkd}}$, scales as $1/\sqrt{{\cal L}}$. 
A best case scenario would take the
current expected limit and scale it linearly with the luminosity. This is equivalent to a statement
that analyses can be developed maintaining the same efficiency and backgrounds as the current
analyses.

For $\tau \rightarrow l \gamma$, there is an irreducible background from 
$\tau \rightarrow l \nu \nu$ for $\epem \rightarrow\tautau$ events accompanied by initial state radiation (ISR).
The ISR
photon can be combined with the lepton $l$ to form a $\tau \rightarrow l \gamma$ candidate that
accidentally falls in the signal region in the 
$M_{\mathrm{LFVD}} - \Delta E$
plane. Such a background is irreducible in the
sense that it arises from an $\epem \rightarrow\tautau$ process with a well measured lepton and $\gamma$ in one 
of the $\tau$
hemispheres. In the existing $\babar$ analyses, these events account for approximately one fifth of
the total residual background. Scaling with this irreducible background only, one would expect an upper
limit for $\BR(\tau \rightarrow l \gamma)$ which ranges between 
$1$ and $2 \; 10^{-8}$ from a complete combined $\babar$ and Belle data
set. 

The situation for the other LFV decays, $\tau \rightarrow 3l$ and $\tau \rightarrow l h h^{\prime}$, 
is even more promising,
since these modes do not suffer from the aforementioned backgrounds from ISR. In this case,
one can project sensitivities assuming $N_{\mathrm{bkd}}$ comparable to backgrounds in existing analyses for
approximately the same efficiencies. These yield expected limits at the $10^{-8}$ level with the
complete $\babar$ and Belle data set.

\section{Conclusions}
Searches for LFV in $\tau$ decays are an optimal hunting ground for BSM physics, which is 
complementary to possible LHC discoveries: observation or non observation of LFV processes 
in the charged sector can significantly constrain theory parameter space.
\begin{figure}%[h]
\includegraphics[width=80mm]{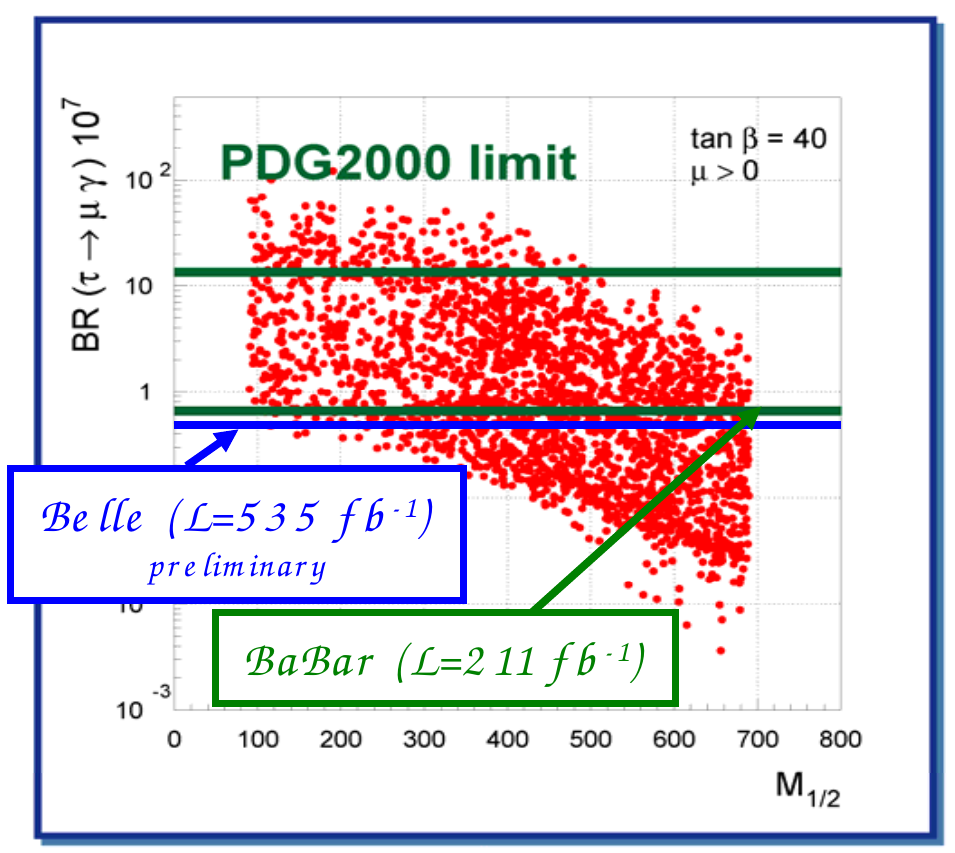}
\caption{$\babar$ and Belle current experimental limits on $\BR(\tau \rightarrow \mu \gamma)$
         compared to a range of theoretical predictions for this branching ratio in a SUSY $SO(10)$ scenario 
         of BSM physics~\cite{MasieroEtAl03}. The best experimental limit available before
         the two asymmetric $B$-factories started to play a role is shown by the uppermost
         dark green horizontal line (``PDG2000 limit'').
         \label{fig:ExpVsTh}}
\end{figure}

$\babar$ and Belle have looked for signals of LVF in many exclusive $\tau$ decay modes and,
although no evidence of LFV has yet been found, limits have pushed into the $10^{-8}$ region 
where parameter space in some BSM scenarios is already being constrained. An example of the constraining power
of the current experimental limits is shown in Fig.~\ref{fig:ExpVsTh}, where the $\babar$
and Belle limits on $\BR(\tau \rightarrow \mu \gamma)$ are superimposed to a scatter plot
of theoretical predictions obtained in a SUSY $SO(10)$ scenario of BSM physics~\cite{MasieroEtAl03}.
Also shown for comparison is the best experimental limit available in the pre-asymmetric-$B$-factory
era.   

It should finally be mentioned that there exist proposals~\cite{SuperB} for Super $B$-factories which
would generate up to a 100 fold increase in the size of the $\tau$ sample compared to those expected
from the existing $B$-factories. If such a facility is built, one will be probing LFV $\tau$ decays 
at the $O(10^{-9} \rightarrow 10^{-10})$ level.

\begin{acknowledgments}
I wish to warmly thank the organizers for the great conference and G. Ridolfi for carefully reading the manuscript.
\end{acknowledgments}

\bigskip % extra skip inserted

\end{document}